\documentclass[twocolumn,preprintnumbers,amsmath,amssymb]{revtex4}

\usepackage{graphicx}
\usepackage{dcolumn}
\usepackage{bm}

\newcommand\bond{$B_{og}$}

\begin{document}

\preprint{APS/123-QED}

\title[]{Frictional weakening induced by cohesion: \\ the numerical case of  cohesive granular failures}

\author{Lydie Staron}
 \email{lydie.staron@sorbonne-universite.fr}
%
 \affiliation{Sorbonne Universit\'e, CNRS - UMR 7190, Institut Jean Le Rond d'Alembert, F-75005 Paris, France}
\author{Laurent Duchemin}
 \affiliation{ESPCI Paris, UMR 7636 - Physique et M\'ecanique des Milieux H\'et\'erogenes, F-75005 Paris, France}
\author{Ana\"\i s Abramian, Pierre-Yves Lagr\'ee}%
\affiliation{ Sorbonne Universit\'e, CNRS - UMR 7190, Institut Jean Le Rond d'Alembert, F-75005 Paris, France}%

\begin{abstract}
The failure of 2D numerical cohesive granular steps collapsing under gravity are simulated for a large range of cohesion.  Focussing on the cumulative displacement of the grains, and defining a displacement threshold, we establish a sensible criterion for capturing the failure characteristics. We are able to locate the failure in time and to identify  the different stages of the destabilisation. We find that the onset of the failure is delayed by increasing cohesion, but its duration becomes shorter.  Defining a narrow displacement interval, a well-defined shear band  revealing the failure comes out. 
Solving the equilibrium of the failing block, we are able to make successful predictions for the dependance between failure angle and cohesion, thereby disclosing two distinct frictional behaviour: while friction remains constant at small cohesion, it significantly decreases with cohesive properties at larger cohesion. The results hence reveal two regimes for the behaviour of cohesive granular matter depending on cohesion strength, revealing a cohesion-induced weakening mechanism.
 \end{abstract}

\maketitle

\section{Introduction}
\label{sec:intro}

One enduring difficulty in describing the behaviour of granular media lies in their ability to adapt external sollicitations by changing behaviour, from flowing like a gas to resist shear like a solid \cite{jaeger96}. Adding cohesion between the grains further obscures the picture: clogging in flows and size-dependent stability threshold  mix up  with effective viscosity and material properties in a way still to be clarified. Because cohesive granular materials are causing much trouble in manufacturing techniques, significant work has  been carried out in the engineering community  to describe the various behaviour of cohesive material and characterise their properties \cite{rumpf70,schubert84,jenike87,pierrat98,muzzio03,emery09,liu21}. Powders are mostly involved, namely very fine grains strongly sticking to each other due essentially to van der Waals forces.  More academic considerations have also prompted numerous works \cite{richefeu06,badetti18,deboeuf21}. Because the exigence for measurable well-constrained quantities often means larger grains sticking together through capillary forces, this generally implies that weaker cohesive forces only are accessible.  Recently,  the trade-off between cohesion control and cohesion strength has been mitigated by the conception of a sticky polymeric coating, thus opening the way to more quantitative measurements at both low and high cohesion \cite{gans20}.  In this context, discrete numerical simulations can be of great help. Although adding cohesive forces in the contact model may actually prove non-trivial \cite{mandal20,staron21}, consistent simulations allow for probing systems behaviour over a large range of parameters \cite{rognon08,berger15,vo20a,vo20b,abramian20,mandal21}.\\
\indent In this contribution, we are interested in characterising the failing behaviour of cohesive columns of varying cohesive properties. While most works on granular columns, including the collapse of cohesive material, concentrate on the run-out behaviour or the deposit shape after the collapse \cite{meriaux08, artoni13,santomaso18,bougoin19,li21,abramian21}, the present work focuses on the first instant of the failure. The objective is twofold. First, we are interested in uncovering a robust criteria for identifying the signature of the failing event and the emergence of a shear band. We then examine how varying the cohesive properties of the contact changes the main characteristics of the failure.  We observe that increasing cohesion delays the instant of the failure and shortens its duration. Interestingly, the orientation of the failure is found to behave non-monotonously with cohesion.\\
 In a second step, we discuss the  implications of the results for the internal frictional properties of the material through analysing the equilibrium of  a cohesive column, with friction and cohesion as sole ingredients. The successful predictions of the model coincides with the existence of a bifurcation in the behaviour of cohesive granular material, defining two distinct frictional behaviours. While weakly cohesive material exhibit a constant internal  friction, larger cohesion involve the significant decrease of friction with cohesive properties, revealing a weakening mechanism.\\
The numerical method and set-up are presented in section \ref{sec:num}. Section \ref{sec:bond30} details the methodology to identify the failure in space and  time for one specific value of the cohesion. The effect of varying the cohesion strength  on the characteristics of the failure are examined in section \ref{sec:allbond}. Implications for the internal friction properties of the material are then discussed in section \ref{sec:discu}.

\section{The numerical cohesive collapse}
\label{sec:num}

\begin{figure}[h]
\begin{minipage}{0.98\linewidth}
\centering

\includegraphics[width=0.8\linewidth,clip]{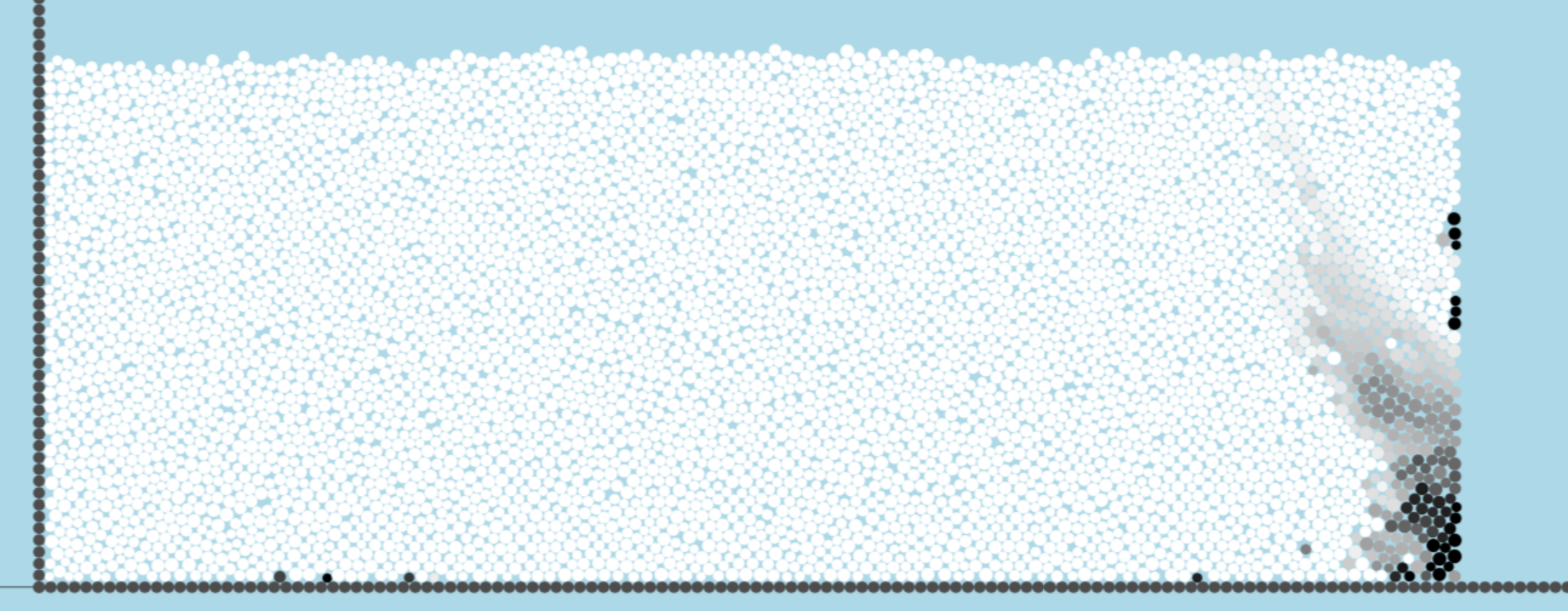}
\includegraphics[width=0.8\linewidth,clip]{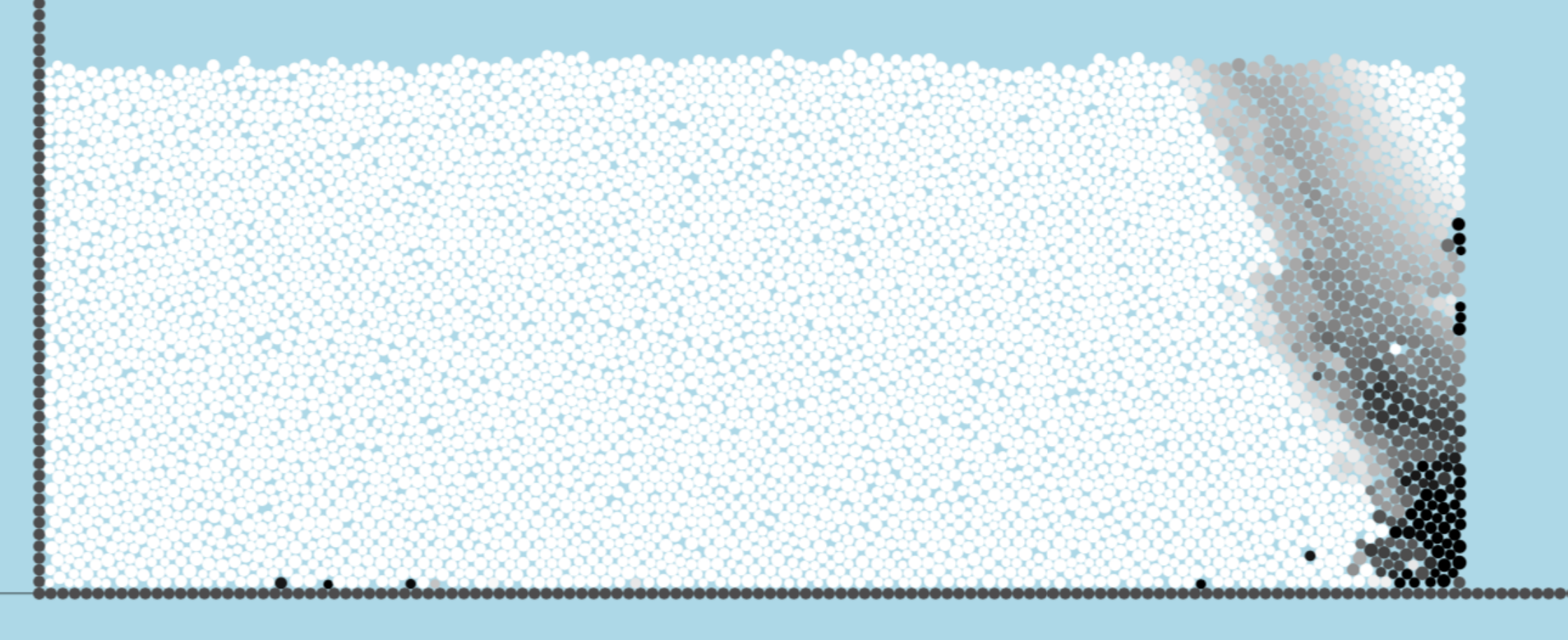}
\includegraphics[width=0.8\linewidth,clip]{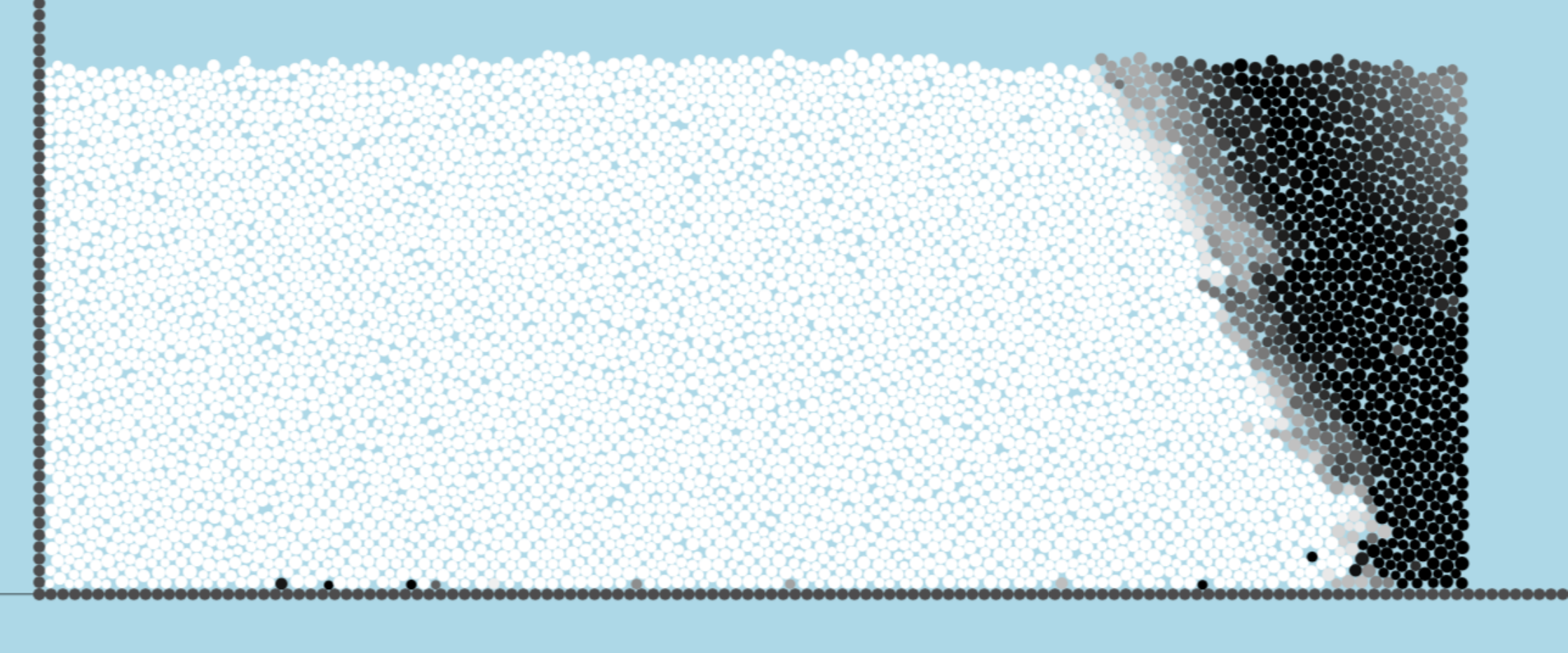}
\includegraphics[width=0.8\linewidth,clip]{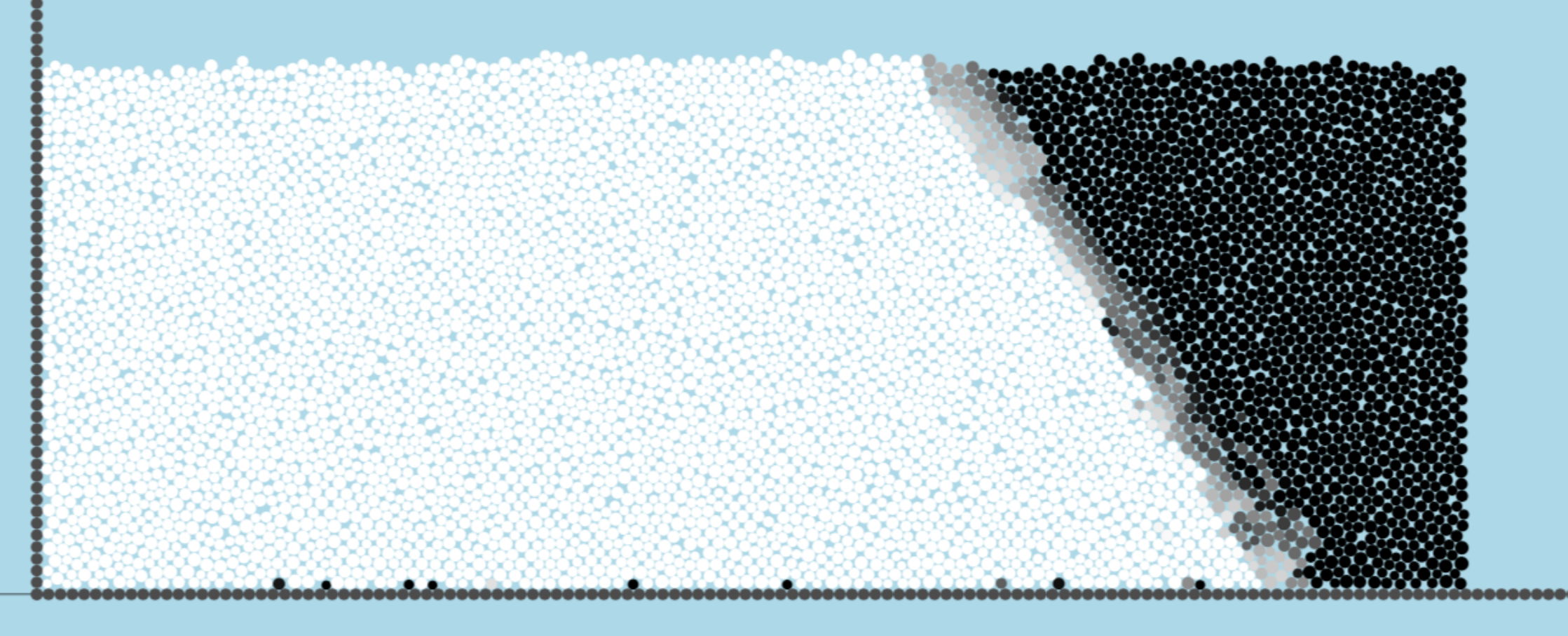}
\includegraphics[width=0.8\linewidth,clip]{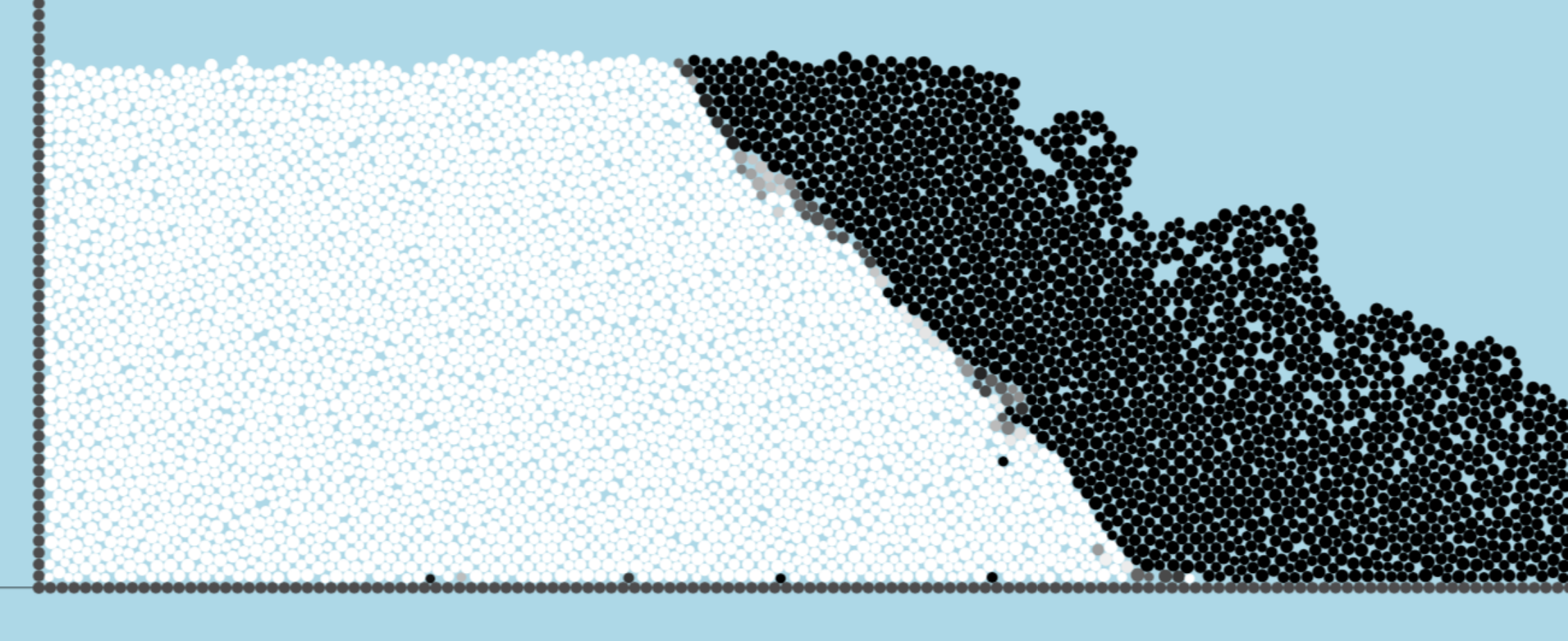}
\caption{Successive snapshots of the initiation of the failure of a cohesive granular step. Black colour coincides with a cumulative displacement  $ >1.4 r_{th} $,  while white colour with a cumulative displacement  $ < r_{th} $, with $r_{th}=0.1d$. The grey colour scale is linear in this interval. Time shown are, from top to bottom, $t/T= 0.10$, $0.11$, $0.12$, $0.17$ and $t/T= \infty$, $T = \sqrt{H/g}=0.1515 \: \text{s}$. (\bond$= 30$).}
\label{fig:flow}       
\end{minipage}
\end{figure}

A  Contact Dynamics algorithm was applied to simulate simple small two-dimensional cohesive systems \cite{kadau02,radjai09}, namely squat columns made of 5572 grains, and left to fail and spread onto a horizontal plane made rough by gluing grains on it. An illustration of the system and its evolution is shown in Figure \ref{fig:flow}. Because the present work is interested in the failure rather than the ensuing spreading, we actually focus on the first instants of the evolution, recording the system state every $\Delta t = 10^{-3}\text{s}$, with a computational time step  $dt= 2.10^{-4}$ s.  The data analysed in the following essentially span the states shown in the  first four pictures of Figure \ref{fig:flow}.  \\
The columns are generated by deposition under gravity of disks with a diameter randomly chosen in the interval $[0.004\text{m}; 0.006\text{m}]$, so that the mean diameter is $d = 0.005 \text{m}$. The random function assigning the sequence of diameters to grains allows for the generation of  fundamentally different, independent initial states in terms of grains and contacts arrangement, yet with an identical size and identical macroscopic dimensions.  The columns thus generated have a height $H \simeq 45 {d}$ and a width $R \simeq 120{d}$. We define the characteristic time $T = \sqrt{H/g}$, where $g=9.81 \text{m.s}^{-2}$. The columns are bounded on the left hand side by a rigid vertical wall, and are free to spread on the right hand side. The dimensions of the system, making it a squat step rather than a column, are thus chosen to ensure that the failure is far enough from the wall to be unaffected by its presence. \\
Each contact is made cohesive through the introduction of a negative ({\em i.e.} in extension) force threshold $-F_c$ depending on the weight of the grains in contact, and a granular Bond number \cite{nase01,rognon08}:
\begin{equation}
F_{c} = B_{og}\: m_{ij}\:g,
\label{eq:fadh}
\end{equation}
with $ m_{ij} = 2 (\frac{1}{m_i} + \frac{1}{m_j})^{-1}$, and $i$ and $j$ are the two grains involved. The Bond number  \bond$ = F_c/mg $, giving the maximum resistance of contacts regarding the weight of one grain, is thus a sensible quantification of the cohesive properties of the granular packing, although other material properties such as  stiffness and elasticity may actually contribute to the effective cohesion in dynamical cases \cite{mandal20}. Note however that contact dynamics consider perfectly rigid grains, thereby disposing of the contact stiffness, and that the present study focuses on the quasi-static dynamics. \\
\indent The Bond number was successively set to \bond$ = 5$, $10$, $20$, $30$, $40$ and $50$. In addition, the non-cohesive case \bond$ = 0$ is also considered. For each value of the Bond number, 11 independent simulations were performed, namely 11 initial states were generated and left to fall, in order to assess the robustness of the observed behaviour. The column height $H \simeq 45 d$ of the systems and the combination of cohesion explored  coincides with  unstable states \cite{abramian20}. If we suppose that the yielding height $H_y \simeq 0.5 B_{ond}$, as is observed for similar systems in \cite{abramian20}, the simulations discussed here satisfy from $H/H_y \simeq 2$ (for \bond$=50$) to $H/H_y \simeq 18$ (for \bond$=5$).  \\
The grains in contact also interact through Coulombic friction, involving the microscopic coefficient of friction $\mu_m=0.2$ (not varied in this paper), and through inelastic collisions. Their volume density is $\rho = 0.1\: \text{kg.m}^{-2}$.

\section{Characterizing the failure: the case of $\boldsymbol{B_{og}=30}$}
\label{sec:bond30}

In this section, we explore simple tools to identify and characterise the failure of the columns. A single value \bond$=30$ of the cohesive force threshold is thus considered.

\subsection{Defining a displacement threshold}

\begin{figure}[h]
\begin{minipage}{1\linewidth}
\centerline{\includegraphics[width=0.99\linewidth]{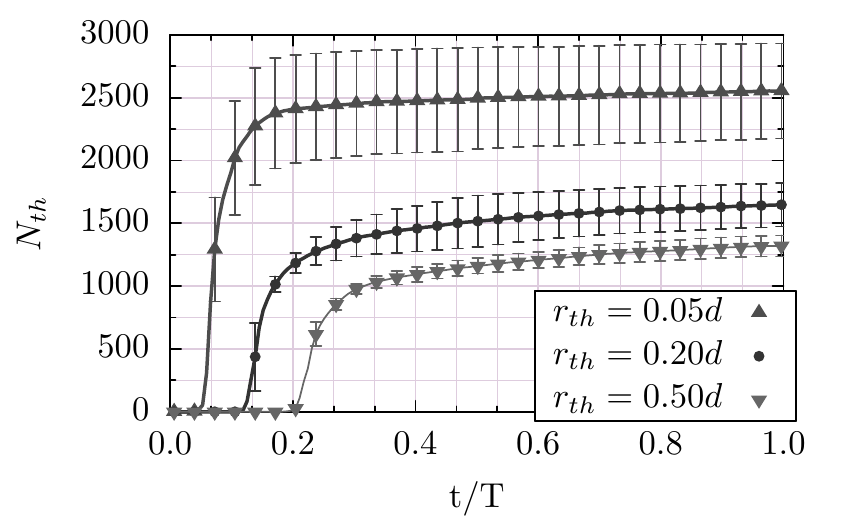}}
\caption{Number of grains $N_{\text{th}}$ whose cumulative displacement exceeds the displacement threshold $r_{\text{th}}$ in the course of time, for three values of  $r_{\text{th}}$, averaged over 11 independant simulations (\bond$=30$). The error bars show the corresponding standard deviation.}
\label{fig:NthvsRth}
\end{minipage}
\end{figure}

The easiest way to characterise the occurence of  the failure of a column/step is certainly to track the displacement of the grains, without presuming the location of the displacement, nor its orientation.  One needs therefor to define a displacement threshold (denoted $r_{\text{th}}$  in the following) to identify those grains which have moved from those which will be considered immobile.\\
 To help making a sound choice, we consider three values for  the displacement threshold:  $r_{\text{th}} = 0.05 d$,  $r_{\text{th}} = 0.20 d$, and $r_{\text{th}} = 0.50 d$.  We note $\Delta r_i$ the cumulative displacement of each grain $i$, and count for each value of $r_{\text th}$, the number of grains $N_{\text{th}}$ whose cumulative displacement  $\Delta r_i$ exceeds $r_{\text{th}}$. The behaviour of $N_{\text{th}}$  in the course of time is displayed in Figure \ref{fig:NthvsRth}  averaged over 11 independent simulations (with the error bars showing the corresponding standard deviation). We observe, for the 3 values of $r_{\text{th}}$, a sharp step-like evolution, with a distinct quick increase, which we identify as the onset of stability loss, namely the occurrence of the failure. This is followed by a rapid saturation which coincides with the flow of detached material  running away, with no significant number of additional grains further displaced beyond the value of the threshold. The different sequences of the stability loss thus remain the same irrespective of the choice of a value for $r_{\text{th}}$.  \\
 The effect of the choice of $r_{\text{th}}$  shows however in  the time at which the sudden jump of $N_{\text{th}}$  occurs and its saturation value. Expectedly, the smaller the displacement threshold, the sooner the grains reach this value, and the larger  the number of grains which eventually overpass the threshold. However, small displacement thresholds induce large error bars, while they nearly vanish for large displacement threshold. Moreover, small values of  $r_{\text{th}}$ also probe diffuse motion of grains in the bulk, far enough from the breaking zone so that filtering them out of the failure event is uneasy and might induce biased choices. On the other hand,  large values of  $r_{\text{th}}$ mean that you miss out the early stages of the failure, with a risk of probing the erosion following the failure, rather than the failure itself.  A trade-off value between these two faulty limits was found to be $r_{\text{th}}=0.20 d$, which we use in  the rest of the paper to characterise the failure dynamics.

\subsection{Locating the failure in time}
\label{sec:trackintime}

\begin{figure}[h]
\begin{minipage}{1\linewidth}
\centerline{\includegraphics[width=0.98\linewidth]{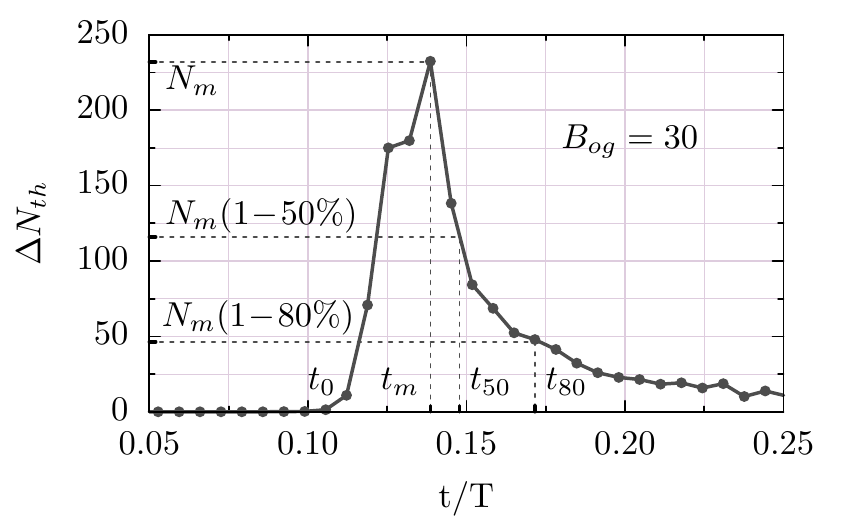}}
\caption{Variations of the number of grains  having overpassed the displacement threshold $r_{th}= 0.20 \text{d}$ in the course of time ($T=\sqrt{H/g}$),  in the case of a cohesion \bond$ = 30$.   }
\label{fig:DNth-Bond30}
\end{minipage}
\end{figure}
     
\begin{figure}[h]
\begin{minipage}{1\linewidth}
\centerline{\includegraphics[width=0.98\linewidth]{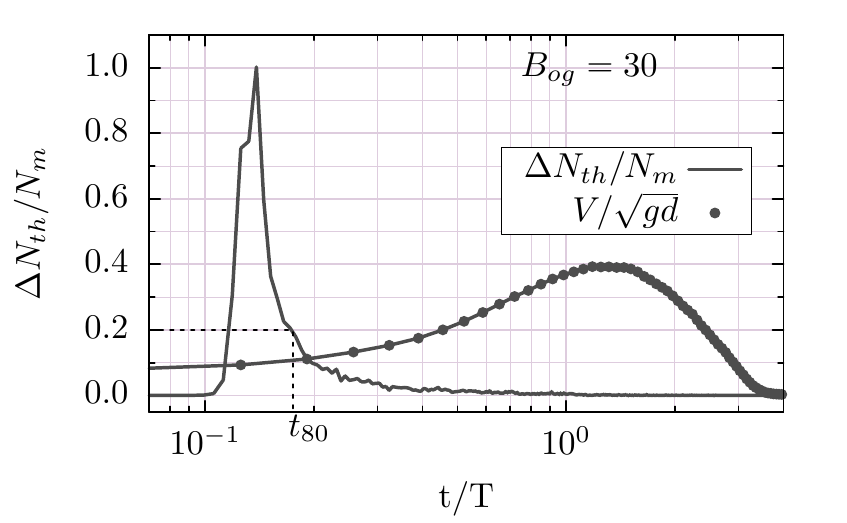}}
\caption{Variations of the number of grains  having overpassed the displacement threshold $\Delta N_{th}$  and mean grains normalised velocity  $V/\sqrt{gd}$ in the course of time, in the case of  a cohesion \bond$ = 30$.   }
\label{fig:Nth-t20-vs-Vmean}
\end{minipage}
\end{figure}

\begin{figure}[h]
\begin{minipage}{1\linewidth}
\centerline{\includegraphics[width=0.98\linewidth]{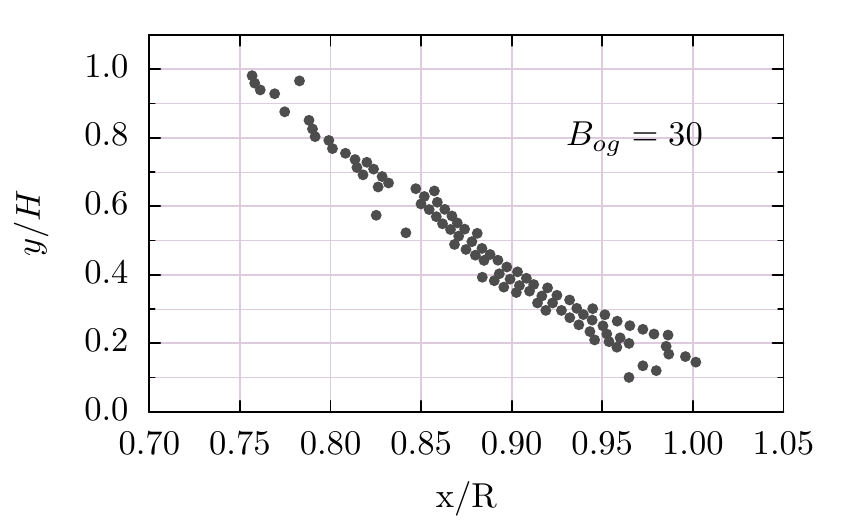}}
\caption{Position of the grains whose cumulative displacement $\Delta r_i$ falls in the interval $[r_{th}: r_{th}(1+10\%)]$ at time $t_{80}$ for one example simulation for \bond$=30$. ($R$ and $H$ are respectively the width and the height of the granular step). }
\label{fig:FailureLine-zC030}
\end{minipage}
\end{figure}

\begin{figure}[h]
\begin{minipage}{0.98\linewidth}
\centering
\includegraphics[width=0.8\linewidth,clip]{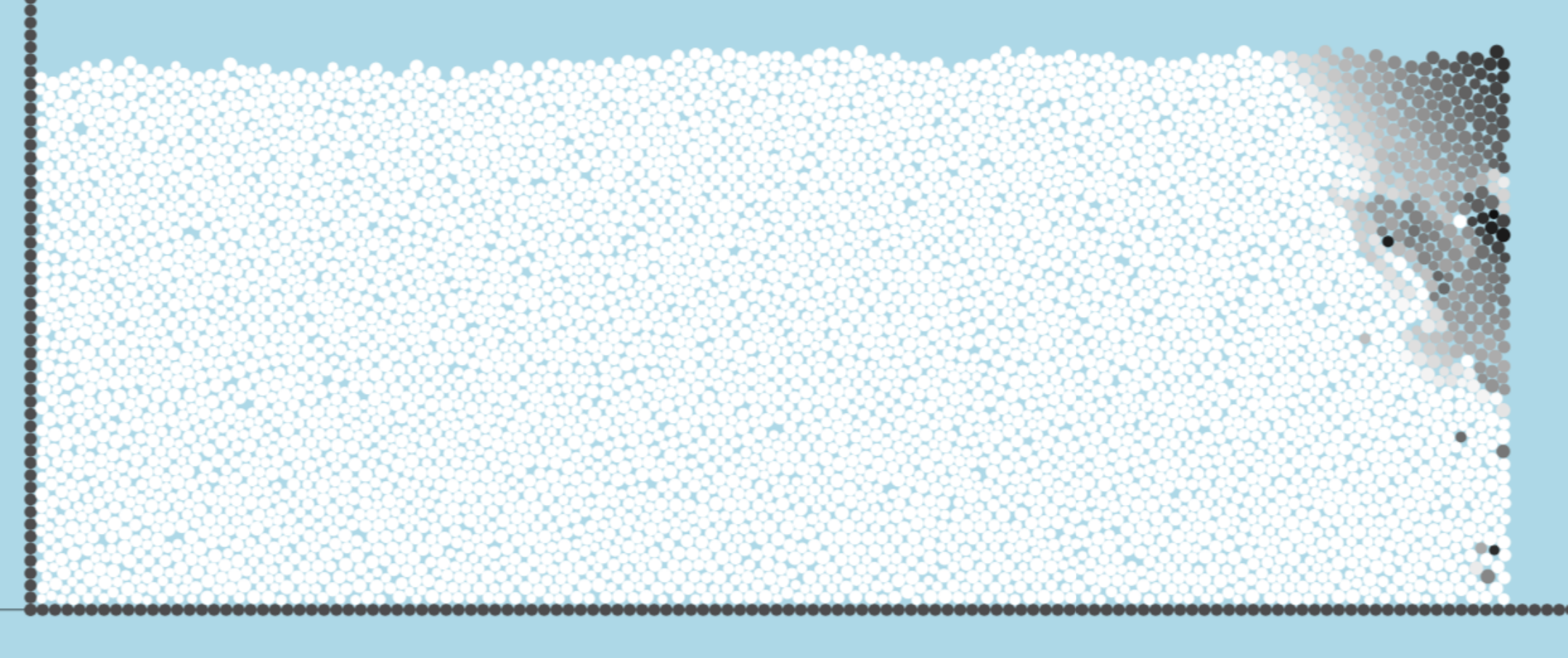}
\includegraphics[width=0.8\linewidth,clip]{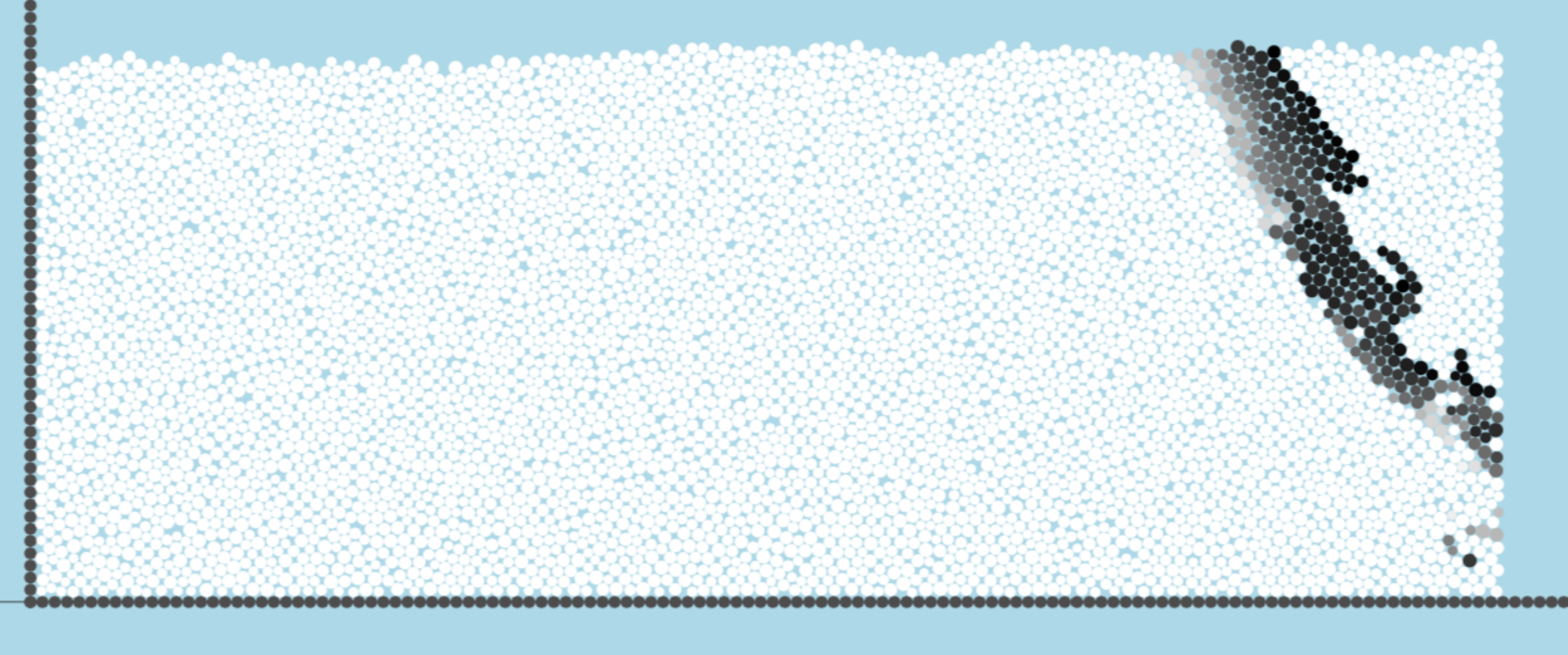}
\includegraphics[width=0.8\linewidth,clip]{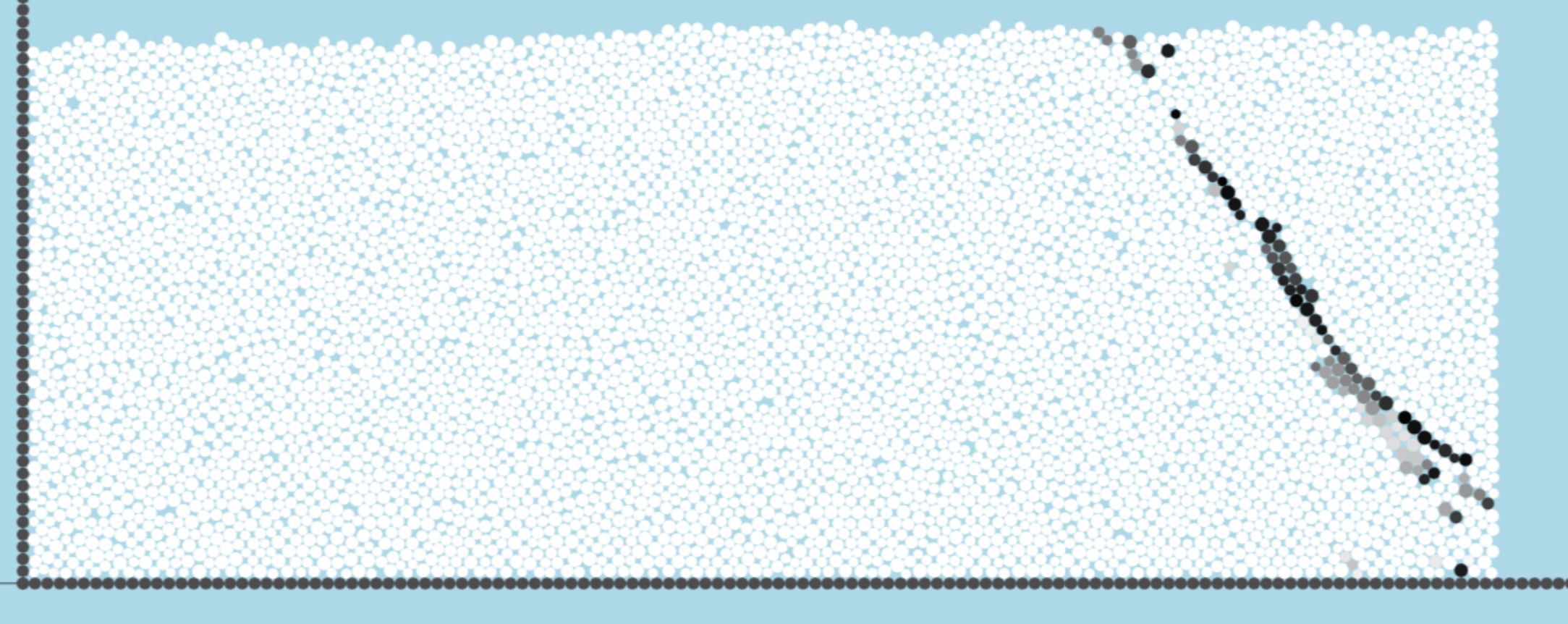}
\caption{Successive snapshots of the failure of an example cohesive granular step with \bond$ = 30$.  White colour coincides with a cumulative displacement  $\Delta r_i < r_{th} $ and $\Delta r_i > r_{th} (1+10\%) $, with $r_{th}=0.20d$. The grey colour scale is linear in this interval, and shows grains such that $\Delta r_i \in [r_{th}: r_{th}(1+10\%)]$. Time shown are, from top to bottom, $t_{m}$,  $t_{50}$ and $t_{80}$ .}
\label{ImageFailure-Bd30}       
\end{minipage}
\end{figure}

     The smooth evolution of $N_{\text{th}}$ allows us to identify the onset of the stability loss, but not exactly the occurrence of a well-defined failure. Yet, in order to compare simulations with different cohesion, we need a robust criteria, easy to implement, easily “portable" from one simulation to another, and physically meaningful. \\
\indent A natural choice is to detect an inflection point in the evolution of  $N_{\text{th}}$, simply plotting $$\Delta N_{\text{th}}= N_{\text{th}}(t+\Delta t) - N_{\text{th}}(t)$$ as a function of time, namely the instantaneous number of grains passing the displacement threshold $r_{th}$. For illustration, $\Delta N_{\text{th}}$ is computed, with $r_{\text{th}}=0.20 d$, for the value of $N_{\text{th}}$ shown in  Figure \ref{fig:NthvsRth}, namely averaged over all runs; the result is plotted in Figure \ref{fig:DNth-Bond30} for the first moments of the evolution. A pic value -- the inflexion point -- clearly comes out, after a rapid ascent corresponding to a sudden mobilisation, bringing an increasing number of grains beyond the displacement threshold. A slower descent follows, corresponding to more localised  motion involving fewer and fewer new grains. We define the time $t_0$ at which motion is detected ($\Delta N_{th}>0$),  the time $t_{m}$ at which the pic (maximum) value $N_{m}$ is reached, and the times $t_z$ at which the instantaneous number of grains reaching the displacement threshold has fallen of $z\%$ of its pic value, {\em i.e.}  for $\Delta N_{th}= N_m (1-z/100)$.  \\
 Obviously, $t_0$, $t_{m}$ and all $t_n$ are specific for each simulation, and are determined computing $\Delta N_{\text{th}}$ for each simulation rather than averaged over all runs, as done in Figure \ref{fig:NthvsRth} for illustration. We find that they take well-defined values. For instance, in the case $B_{ond}=30$ discussed in this section, analysing the 11 independent simulations gives $t_0/T \simeq 0.111 \pm 8.10^{-3}$ and $t_{m}/T \simeq 0.140  \pm 7.10^{-3}$.  Looking at the chronology of the decrease from the pic value, we find $t_{50}/T \simeq 0.147 \pm 8.10^{-3}$, $t_{80}/T \simeq 0.157  \pm 9.10^{-3} $ and $t_{90}/T \simeq 0.184   \pm  14.10^{-3} $. \\

  \indent Because determining $t_z$ is not straightforward for small values of $z$ ({\em i.e.} shortly after the maximum, when the fall of $\Delta N_{\text{th}}$ is very stiff) and will be somehow blurred by the value of the numerical computational time step, we prefer to focus on a latter stage of the evolution, coinciding with a larger fall of $\Delta N_{\text{th}}$, namely $t_{80}$. The graph in Figure \ref{fig:DNth-Bond30} may give the feeling that $t_{80}$ is already at the end tail of the failure process; and certainly some grains mobilised at  $t_{80}$ are responding to the beginning of the propagation of the failure  through erosion rather than being a picture of its very first instants. However, plotting $\Delta N_{\text{th}}/N_{m}$ together with the mean grain velocity $V/\sqrt{gd}$  (Figure \ref{fig:Nth-t20-vs-Vmean}) shows that $t_{80}$ is still in the early stage of the failure, so that we do not expect the effect of failure propagation to be dominating. In addition, doing so we exclude the early diffuse motion of the grains. \\
 We observe in Figure \ref{fig:Nth-t20-vs-Vmean} that the number $\Delta N_{\text{th}}$ of new grains displaced beyond the threshold, soon vanishes, leaving the dynamical part of the failure, where the mean velocity $V$ becomes significant, to the population of grains already detached. In other words, there is a clear dichotomy between the spreading of the failing material and the (nearly) static grains remaining part of the packing.

\subsection{Locating the failure in space}
\label{sub:failinspace}

\begin{figure}[h]
\begin{minipage}{1\linewidth}
\centerline{\includegraphics[width=0.98\linewidth]{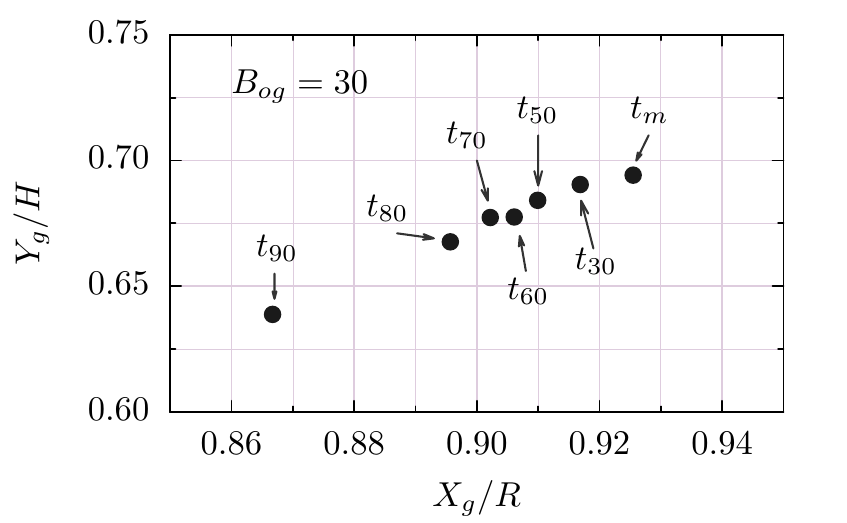}}
\caption{Position of the centre of mass of the grains such that  $\Delta r_i \ge r_{th}$ at the different stages of the failure, from instant $t_m$ to instant $t_{90}$ (see text for definition of instants $t_z$), for a cohesion \bond$= 30$.}
\label{fig:Propag-XgYg30}
\end{minipage}
\end{figure}

Now that we have settled that  the failure is best identified at the time $t_{80}$ when the number of grains being displaced beyond the threshold $r_{th}$ has fallen of $80\%$ of its maximum value, we can track its location in the packing. \\
\indent We can hence look at the position of the grains at time $t_{80}$ whose cumulative displacement $\Delta r_i$ falls in the interval $[r_{th}: r_{th}(1+10\%)]$. Doing so, we ignore the grains which have not yet been significantly displaced by the instability, and we filter out the grains who have been displaced 10$\%$ beyond the threshold, thereby excluding those moving away.
The result is displayed in Figure \ref{fig:FailureLine-zC030} for one given example simulation. We observe a well-defined shear zone, which can be approximated by an affine function; the present example shows a slope  $-1.33$, namely a failure oriented at $\alpha = 53^\circ$ with the horizontal, according to a linear approximation  with an asymptotic standard error of $2.2\%$. Another picture of the failure is given by coloring grains such that  $\Delta r_i \in [r_{th}: r_{th}(1+10\%)]$ while less mobile grains ($\Delta r_i <r_{th}$) and flowing-away grains ($\Delta r_i > r_{th}(1+10\%))$ are left in white, as shown in Figure~\ref{ImageFailure-Bd30} for $t_{m}$, $t_{50}$ and $t_{80}$.\\
\indent Following the steps described above, we analyse all 11 independent simulations with $B_{ond} = 30$. We find a  failure oriented at $\alpha = 56.8 \pm  3.7$ degrees, with linear regressions with a mean asymptotic standard error of $2.36 \%$, and a maximum  asymptotic standard error less than $4.5 \%$.\\
\indent Although we focus on the time $t_{80}$ to characterise the failure, the other instants of the decrease can yield interesting  information on the propagation of the destabilisation. For instance, starting from the instant of the pic value $t_{m}$, we consider the position of the grains who have moved beyond the threshold $r_{th}$ in the successive time intervals $[t_m,t_z]$, for each independent simulation. We then  compute the position of the centre of mass of the displaced grains in the interval $[t_m,t_z]$ for each $t_z$. We then average over all simulations. This gives us an averaged integrated picture of the propagation of the instability in the course of time, displayed in Figure \ref{fig:Propag-XgYg30}.

\section{Effect of cohesion on failure anatomy: $\boldsymbol{B_{og} \!=\!0}$ to $\boldsymbol{B_{og} \!=\!50}$}
\label{sec:allbond}

In the section above, we have discussed a criteria to identify the signature of the occurrence of a failure, in time and in space. Therefor, we have tracked the motion of the grains compared to a displacement threshold. This was done for one value of the cohesion $B_{og}=30$. We now apply the same criteria for the same 11 independent columns, but varying the cohesion, through varying the $B_{og}$ number. The following values are now considered:  $B_{og}=0$, $5$, $10$, $20$, $40$ and $50$. We  then discuss the influence of the cohesion on the characteristics of the failure: the time of occurrence, its duration and amplitude, its propagation and its orientation. \\

\begin{figure}[h]
\begin{minipage}{1\linewidth}
\centerline{\includegraphics[width=0.98\linewidth]{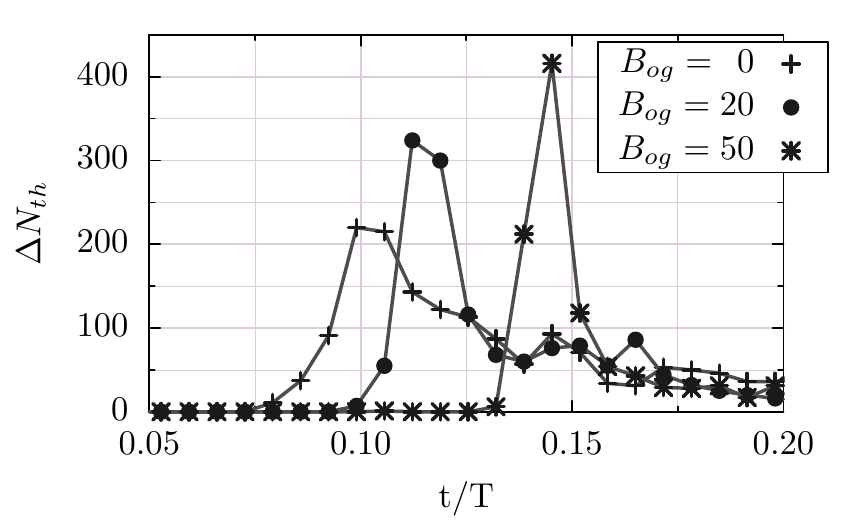}}
\caption{Example of the instantaneous number $\Delta N_{\text{th}}$ of new grains displaced beyond the threshold $r_{th}=0.20d$ in the course of time  for $B_{og}=0$, $B_{og}=20$ and $B_{og}=50$, shown for one simulation.  }
\label{fig:DNth-vs-Bd}
\end{minipage}
\end{figure}

\subsection{A latter shorter event }

\begin{figure}[h]
\begin{minipage}{0.98\linewidth}
{\includegraphics[width=0.97\linewidth]{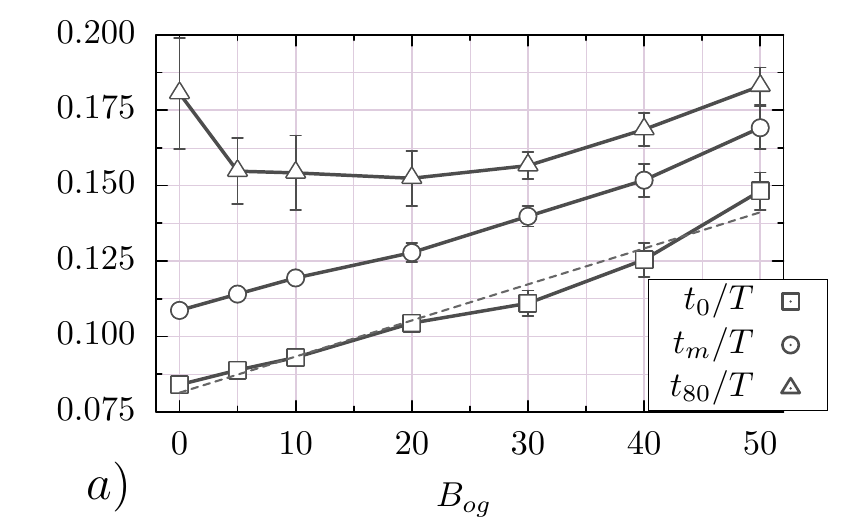}}
{\includegraphics[width=0.98\linewidth]{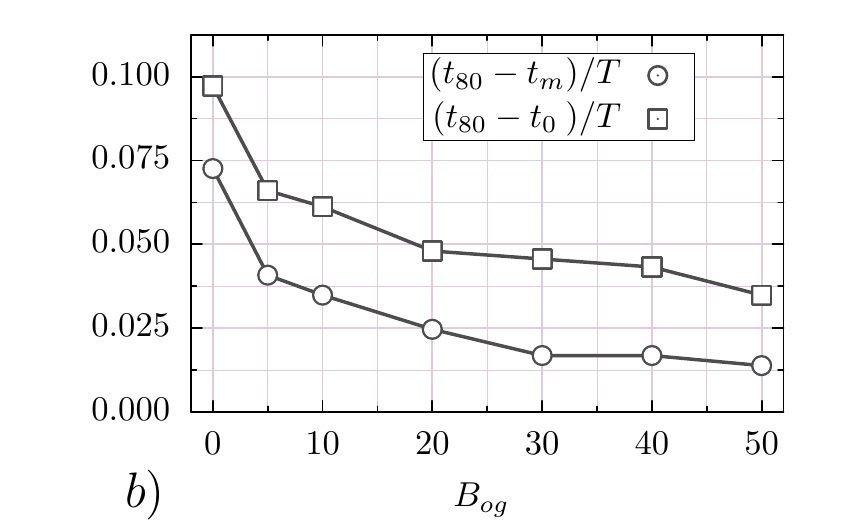}}
\caption{Progression of the failure: a) instants $t_0$, $t_m$ and $t_{80}$ as define in Figure \ref{fig:DNth-Bond30} and b) evaluation of the duration of the failure $(t_{80} - t_m)$  and $(t_{80} - t_0)$  as a function of the cohesive Bond number $B_{og}$. ($T=\sqrt{H/g}$)}
\label{fig:Time-vs-Bond}
\end{minipage}
\end{figure}

The instantaneous number $\Delta N_{\text{th}}$ of new grains displaced beyond the threshold $r_{th}$ in the course of time is shown in Figure \ref{fig:DNth-vs-Bd} for $B_{og}=0$, $B_{og}=20$ and $B_{og}=50$ for one given simulation.  Each curve bears the signature of the onset of a failure, as discussed in section \ref{sec:trackintime}. These informations are each for one specific failure event (one simulation at a given $B_{ond}$), yet they contribute to understanding the effect of cohesion on failure. In particular, Figure \ref{fig:DNth-vs-Bd} plainly reveals the dependence between the instant of the destabilisation, the number of grains displaced, and the cohesion.\\
\indent We first analyse the chronology of each individual failure, writing down the time of initiation $t_0$, the time $t_m$ of the pic value  in terms of number of grains involved, and the assumed instant of the failure $t_{80}$, as they are defined in section \ref{sec:trackintime}. This is done for each of the 11 independent simulations for each value of the $B_{og}$ number, from $0$ et $50$.  The results are displayed in Figure \ref{fig:Time-vs-Bond}. \\
 %
 
\indent We observe that cohesion induces a delayed failure compared to non-cohesive cases; this phenomena is observed experimentally in \cite{artoni13}.  It is not surprising if we consider that the failure results from an avalanche of bonds opening so that global equilibrium is compromised: the stronger the bonds, the larger the number of opening contacts necessary to entail a failure, the longer the time to evolve towards instability. \\
 The time $t_0$ of the onset of the destabilisation  is found to evolve roughly linearly with the  $B_{og}$ number (Figure \ref{fig:Time-vs-Bond}-a).  The associated  error bars (representing the standard deviation over the 11 independent simulations run for each value of $B_{og}$) are small, revealing a well-defined time-scale for the initiation of the destabilisation. Interestingly, while the error bars for $t_0$ are very small for small cohesion, they increase with the $B_{og}$ number: cohesive systems are more “noisy“ in terms of onset time than non-cohesive ones .\\
The instant $t_m$, coinciding with the maximum value of $\Delta N_{\text{th}}$, follows the same trend as $t_0$. By contrast, the instant $t_{80}$ shows a non-monotonous evolution. In addition, the corresponding error bars soar, especially for low cohesion. If we define the duration of the failure as $t_{80}-t_m$ (or alternatively, $t_{80}-t_0$), we find that large cohesions induce shorter failure, while low cohesion induces significantly longer events (Figure \ref{fig:Time-vs-Bond}-b).


\subsection{ A slower failure of varying orientation}

\begin{figure}[h]
\begin{minipage}{1\linewidth}
\centerline{\includegraphics[width=0.98\linewidth]{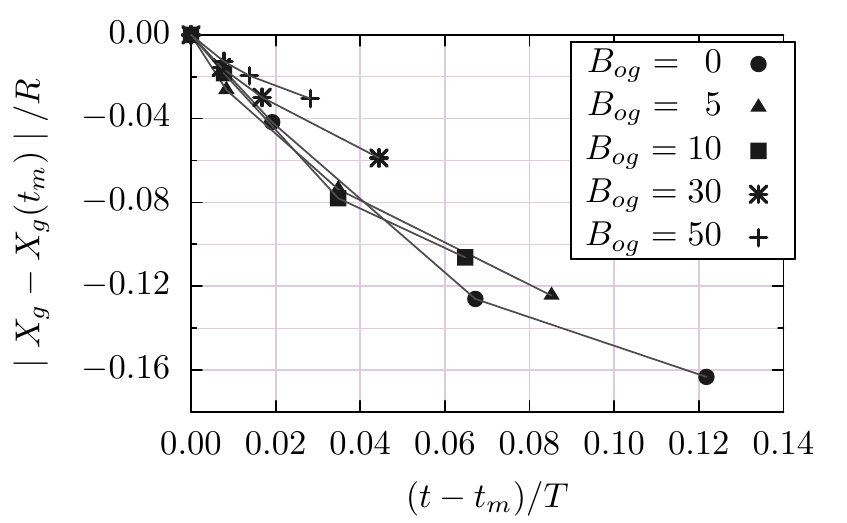}}
\caption{Horizontal position of the centre of mass $X_g$ of the displaced grains such that  $\Delta r_i \ge r_{th}$  with respect with its position at time $t_m$ as a function of the normalised time $(t-t_m)/T$, for $B_{og}=0, 5, 10, 30$ and $50$. For each value of \bond, the three points defining the failure propagation correspond to the times $t_{50}$, $t_{80}$ and $t_{90}$. ($T=\sqrt{H/g}$).}
\label{fig:Propag-Xg-All}
\end{minipage}
\end{figure}

\indent If we focus on the successive time intervals $[t_m,t_z]$ for increasing values of the cohesive $B_{og}$ number, we can derive a mean picture of failure propagation  with time by plotting the horizontal position of the centre of mass $X_g$  of the displaced grains such that  $\Delta r_i \ge r_{th}$  in the time interval $[t_m,t_z]$ with respect with its position at time $t_m$. This is displayed in Figure \ref{fig:Propag-Xg-All} for $t_{50}$, $t_{80}$ and $t_{90}$, and $B_{og}=0, 5, 10, 30$ and $50$. We observe that smaller cohesion coincides with a failure traveling further, at a seemingly greater speed (although the data to not allow for drawing  definite conclusions on failure velocity). \\

\begin{figure}[h]
\begin{minipage}{1\linewidth}
\centerline{\includegraphics[width=0.98\linewidth]{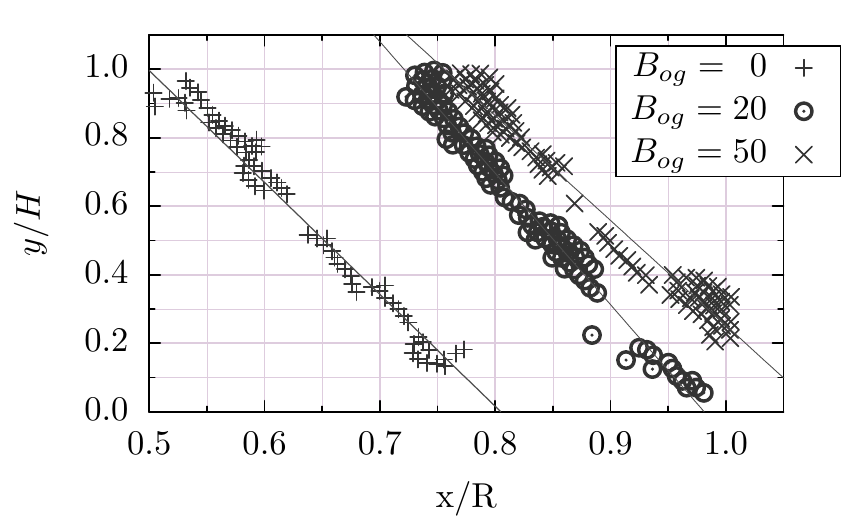}}
\caption{Position of the grains displaced  in the interval  $[r_{th}, r_{th}(1+10\%)]$  at time $t_{80}$,  with the corresponding affine approximation, for $B_{og}=0$, $20$ and $50$, in three example simulations.}
\label{fig:FailureLine-All}
\end{minipage}
\end{figure}

For each simulation, we focus on the position of the grains whose cumulative displacement $\Delta r_i$ falls is the interval $[r_{th}, r_{th}(1+10\%)]$ at  time $t_{80}$.  As done in subsection \ref{sub:failinspace}, we derive the orientation of the failure by approximating the cloud of points by an affine function. Examples of  clouds of points corresponding to the position of the displaced grains  at $t_{80}$ and the corresponding affine approximation are shown in Figure \ref{fig:FailureLine-All} for $B_{og}=0$, $20$ and $50$. We observe failures of various orientations and depth.  \\[5pt]
%
%
\begin{figure}[h]
\begin{minipage}{1\linewidth}
\centerline{\includegraphics[width=0.98\linewidth]{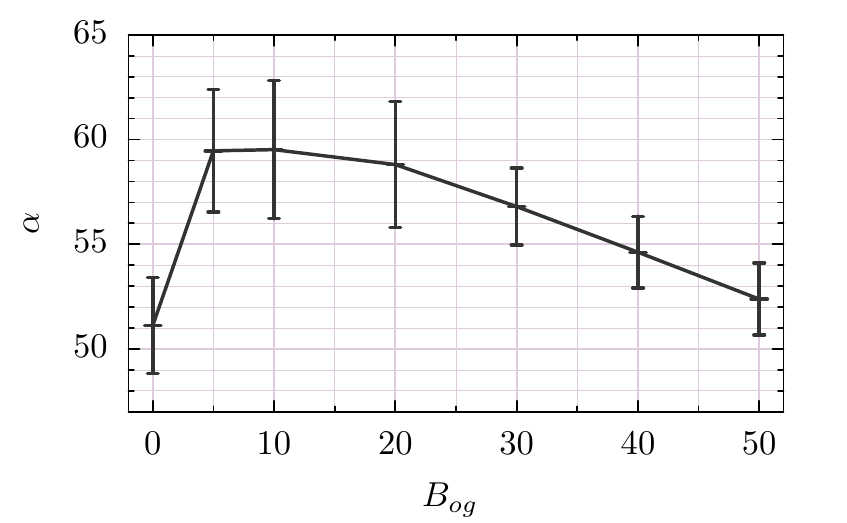}}
\caption{Angle $\alpha$ (in deg) formed by the failure and the horizontal as a function of the cohesive Bond number $B_{og}$. The errors bars show the corresponding standard deviation.}
\label{fig:SlopeAngle-vs-Bond}
\end{minipage}
\end{figure}
\indent Considering all simulations for every value of the Bond number, we can plot the angle $\alpha$ made by the failure line as a function of $B_{og}$ (Figure  \ref{fig:SlopeAngle-vs-Bond}).  The asymptotic standard error associated with the linear approximation varies from $4.3\%$ to $1.8\%$, and decreases consistently when cohesion increases, meaning  a better precision  for larger cohesion (not shown). The error bars associated to the shear slope remain non-negligible, showing a standard deviation of up to $10\%$ for $0 \leq B_{og}  \leq 20$, but about $6\%$ for larger cohesion.\\[1pt]
A  clear trend comes out, showing a non-monotonous behaviour. For small values of $B_{og}$, in the interval $0\leq B_{og}\leq5$, an increasing $B_{og}$ implies a stiffer failure slope. For larger cohesion however,  increasing $B_{og}$  in the interval $5\leq B_{og}\leq50$ leads consistently to  smaller failure slopes. Altogether, the orientation of the shear band varies between $51^\circ$ to $60^\circ$ (Figure \ref{fig:SlopeAngle-vs-Bond}).\\
\indent The shear bands measured here are identified using the same criteria for all Bond number: same threshold for grains displacement, same instant regarding the whole failure event, namely same stage of the failure. We can hence suppose that they convey a comparable information regarding the failure orientation for all values of $B_{og}$. We thus discuss in the next section the evolution of the failure orientation  with cohesion, and more specifically, the implications for the material frictional  properties. In particular, we give evidence of a cohesion-induced weakening mechanism.

%
\section{Discussion on the implications for the internal friction angle}
\label{sec:discu}

\begin{figure}[h]
\begin{minipage}{1\linewidth}
\centerline{\includegraphics[width=0.8\linewidth]{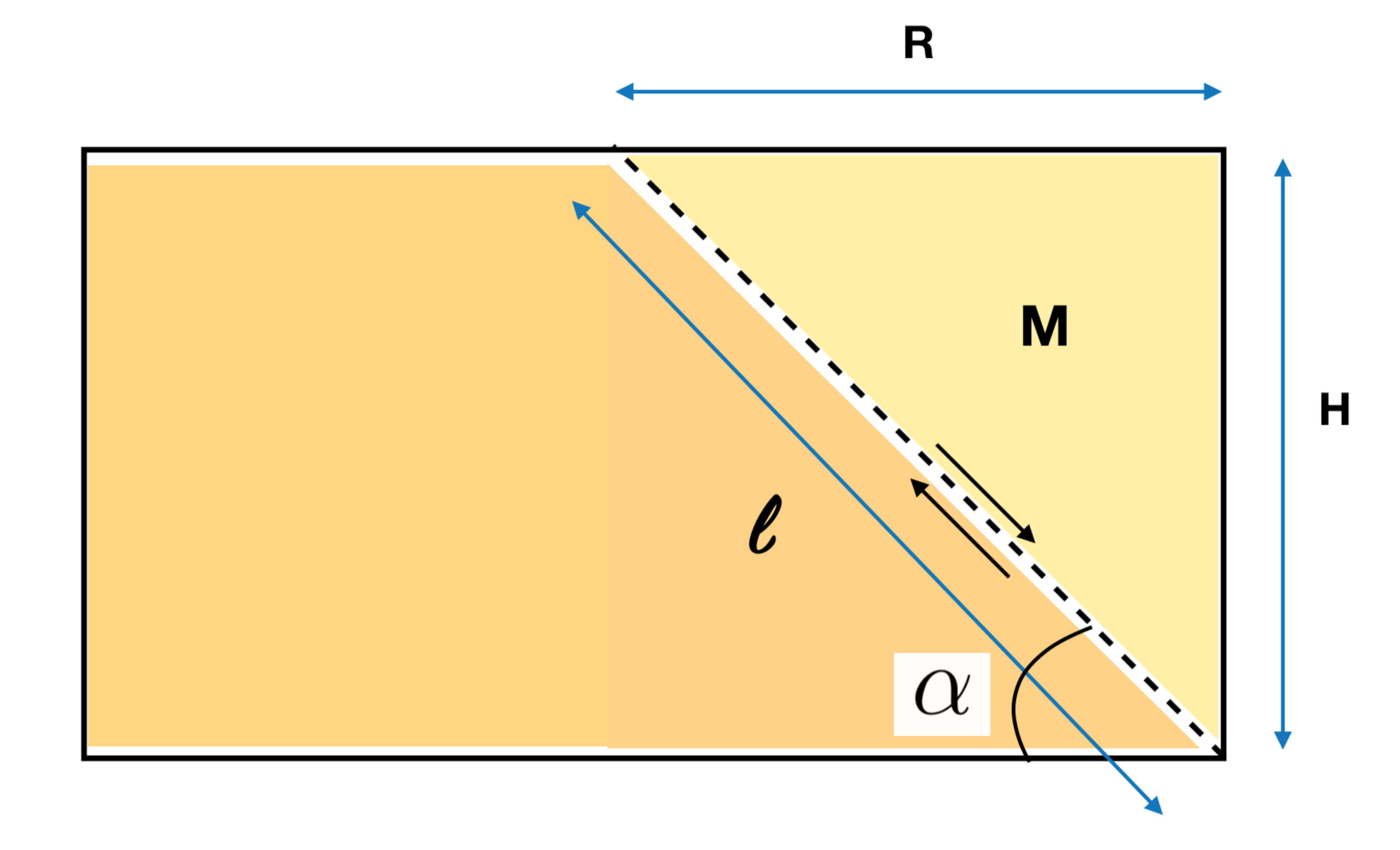}}
\caption{Stability of a cohesive granular step: slip motion of a corner of mass $M$ along the failure plane at incipient failure. $R$ and $H$ are respectively the horizontal extent of the failing corner and the height of the step; $\ell$ is the length of the failure plane.}
\label{fig:StabScheme}
\end{minipage}
\end{figure}

It is generally accepted that cohesion and friction are independent quantities. Indeed, significant variations of the effective coefficient of friction with cohesion were  never observed to our knowledge.
  Experiments of granular beds flowing under gravity or continuous shearing allow for estimating a coefficient of Coulombic friction from the measure of the shear stress and the pressure  \cite{pierrat98,richefeu06,gans20}. The coefficient thus derived is found to be exhibiting a rather constant value.  Since most of these experimental works on cohesive granular matter involve wet grains, namely capillary forces, the cohesion implied is essentially of low to moderate strength. But strongly cohesive  beds do not flow, so that determining  shear stress and  pressure over a range of  cohesion including larger values, to derive an estimation of the friction, is hardly feasible through this method. On the other hand, the angle of repose of a cohesive granular pile increases with cohesion \cite{nase01,samadani01,nowak05, lumay12,gans20}; but while this angle can be easily related to friction properties for non-cohesive material, it is no longer the case for cohesive ones \cite{halsey98}.  An alternative estimation of the frictional properties of the material is thus discussed in the following.\\
Solving the equilibrium  of a cohesive granular step gives us means of approaching the dependance of the internal friction angle on the cohesive properties. Considering the stability  along a plane oriented at an angle $\alpha$ with the horizontal,  supposing a macroscopic cohesion $\tau_c$ , and applying a simple Mohr-Coulomb criterion  \cite{restagno04,abramian20,gans21}, as illustrated in Figure  \ref{fig:StabScheme},  the equilibrium of the upper corner  is broken when
\begin{equation}  
Mg\sin\alpha  =  \mu  Mg\cos\alpha + \tau_c \ell, 
\label{eq:failure1}
\end{equation}
where $\ell$ is the length of the failure plane, and $M$ is the mass of the failing part. If $H$ is the height of the step, and $\rho$ the density of the material, then we have $M~=~\frac12 \rho H^2/\tan \alpha$, and $\ell = H/\sin \alpha$. Equation (\ref{eq:failure1}) can simply be rewritten:
\begin{equation}  
 \tan \alpha  = \mu + \frac{2 \tau_c}{\rho g H} \frac{1}{\cos^2 \alpha}.
\label{eq:failure2}
\end{equation}
The non-cohesive case $\tau_c \!=\!0 $ coincides with the familiar dry granular case $\mu\!\!=\!\!\tan \alpha$, where failure occurs following the free surface  \cite{nedderman92}.  The simulations give a failure angle $\alpha \simeq 1.24 \pm 0.21$ for non-cohesive steps, {\it i.e.}  $\alpha \simeq 51.12^\circ \pm 4.5 $.  Accordingly, equation (\ref{eq:failure2}) would imply $\mu= \tan\alpha \simeq 1.24$. This is a massive figure for granular media in general, and all the more so for circular beads with an inter-particle friction $\mu_c=0.2$. Such values for $\alpha$ are nevertheless observed for experiments of failing granular columns of circular beads  with weak cohesion \cite{gans21}. \\

A way of questioning the behaviour of the internal friction $\mu$ is through minimising the height for which failure occurs, namely finding the condition for which  the height reaches the yield condition $H=H_{y}$.  Equation (\ref{eq:failure1}) can easily be written in the form:
\begin{equation}
H  = \frac{2 \tau_c}{\rho g} \frac{1}{(\cos \alpha \sin \alpha - \mu \cos^2 \alpha)}
\end{equation}
which, considering the internal angle of friction $\phi$ such that $\mu = \tan\phi$, becomes
\begin{equation}
H = \frac{2 \tau_c}{\rho g} \frac{\cos \phi}{ \cos\alpha \sin(\alpha - \phi)}.
\label{eq:Hy}
\end{equation}
The minimum height $H_y$ of a failing step is thus given by minimising the function $1/ \cos\alpha \sin(\alpha - \phi)$, who has a minimum at $\cos(2\alpha - \phi) = 0$, namely for a failure plane such that $\alpha = \pi/4 + \phi/2$. Conversely, the friction angle satisfies $\phi = 2 (\alpha - \pi/4)$ when failure occurs at $H_y$.  \\
\indent  However, the  granular steps simulated here do not have a height H equal to $H_y$.  Moreover, experiments  show that $\alpha$ increases when the column height decreases \cite{gans21}, so that we can suppose that we would observe larger values of the failure angle $\alpha$ for $H\rightarrow H_y$.  For this reason, we cannot discuss quantitatively the internal friction   that can be anticipated from the evolution of $\alpha$ (which would fall in the range  between $16^\circ$ and $32^\circ$).\\
 \indent Nevertheless a trend in the anticipated  evolution of the friction $ \phi =2 (\alpha - \pi/4)$ with the cohesion comes out considering the behaviour of the failure angle $\alpha$ shown in Figure \ref{fig:SlopeAngle-vs-Bond}. For  large values of the cohesion, namely $B_{og} >5$, the data showing $\alpha$ decreasing with \bond,  would imply that friction decreases with cohesion. Interpreting datas using equation (\ref{eq:Hy})  requires however simulating systems close to stability limit, namely with a height very close to the threshold value $H~\simeq~H_y$. \\[5pt]
%

%

\begin{figure}[h]
\begin{minipage}{1\linewidth}
\centerline{\includegraphics[width=0.99\linewidth]{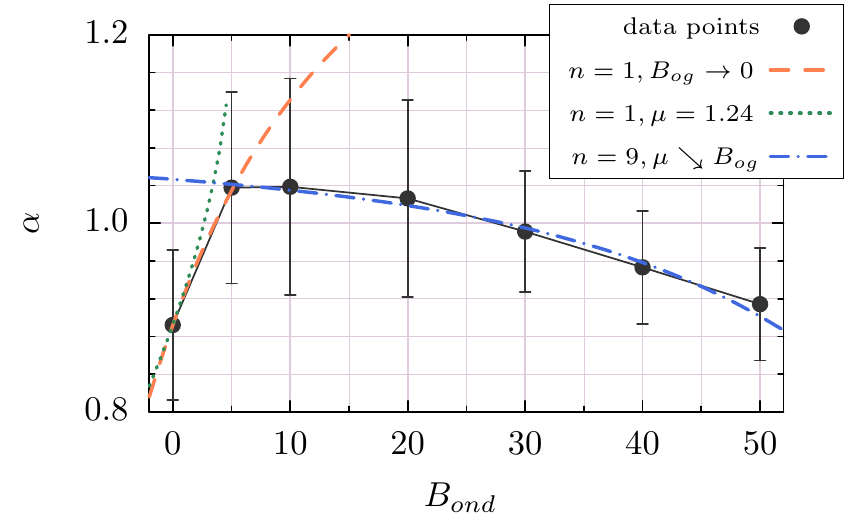}}
\caption{Failure angle $\alpha$ measured in the simulations (full circles) and predictions corresponding to  the approximation for $B_{og}\rightarrow 0$ and $n=1$ (eq. \!(\ref{eq:root2})), to the full solution with a constant friction $\mu = 1.24$ and $n=1$ (eq. \!(\ref{eq:root1})),   and the full solution with a varying friction $\mu = 1.73 - 0.0194 B_{og}$ and $n=9$ (eq. \!(\ref{eq:root1})). }
\label{fig:prediction}
\end{minipage}
\end{figure}
However, more insights can be derived from equation (\ref{eq:failure2}) through simple arithmetics. Equation (\ref{eq:failure2})  can be rewritten 
\begin{equation}  
 0 =    \frac{2 \tau_c}{\rho g H}  \tan^2\alpha - \tan\alpha + \frac{2 \tau_c}{\rho g H} + \mu .
\label{eq:failure3}
\end{equation}
 To introduce the Bond number in this relation, we need  an estimate for the macroscopic cohesive stress $\tau_c$.
 Considering the (linear) surface $S$ over which the adhesion force applies, we merely write $\tau_c = {F_c}/{S}$. In the absence of an indisputable choice for  $S$, we suppose $S=n d$, with $d$ the mean grain diameter, and where $n$ is unknown; we suppose $n\in[1:10]$ \cite{abramian20}. Using the expression of the contact adhesion force threshold $F_c = B_{og}  m g$, we obtain
\begin{equation}  
  \frac{\tau_c}{\rho g} =  \frac{\pi d \times B_{og}}{ 4 n}.
\label{eq:yieldstress} 
\end{equation}
Accordingly, equation (\ref{eq:failure2}) becomes
\begin{equation}  
 \tan \alpha  = \mu + \frac{\pi d B_{og} }{2 n H}  \frac{1}{\cos^2 \alpha}.
\label{eq:failure2bis}
\end{equation}
Replacing ${\tau_c}/{\rho g}$ in equation (\ref{eq:failure3}) gives
\begin{equation}  
0  = \frac{\pi d B_{og}}{2 n H}  X^2 -X +  \frac{\pi d B_{og}}{2 n H} + \mu, 
\label{eq:failure4}
\end{equation}
with $X=\tan \alpha$. Thus, the equilibrium of a cohesive granular step, assuming a constant cohesion $B_{og}$ and a constant friction $\mu$,  is a second degree polynomial relation. Interestingly, its determinant is itself a second degree polynomial function of the $B_{og}$ number:
\begin{equation}
 \Delta = -(\pi d/nH)^2 B_{og}^2 - 2\mu (\pi d/nH) B_{og} +1, 
 \label{eq:determinant}
 \end{equation}
 which allows us to examine the interval of cohesion in which a solution for equation  (\ref{eq:failure4}) exists. We find that  $\Delta$ takes positive values in  the interval $ [B'_c, B_c ]$, with $B'_c$ negative, and $B_c = nH (\sqrt{\mu^2+1} - \mu)/\pi d$ is positive. In other words,  solutions for equation (\ref{eq:failure4}), assuming a constant friction $\mu$, exist  only for a cohesion lower than a critical value $0\leq B_{og}\leq B_{c}$. This would imply that $\mu$ is constant as long as $B_{og}\leq B_{c}$, but a constant friction  is no longer admitted for larger values of cohesion $B_{og}>B_c$. In this domain, the hypothesis of a friction adapting the cohesion $\mu = \mu(B_{og})$ should become an option.\\[5pt]
In the interval $[0,B_c]$, equation (\ref{eq:failure4}) has two roots, one of which diverges for \bond$\rightarrow 0$ (we dismiss this solution). The second root, which does not blow up when  \bond$\rightarrow 0$,  reads 
\begin{equation}
X = \frac{1 - \sqrt{1-\frac{2\pi d B_{og}}{nH} \left( \frac{\pi d B_{og}}{2 nH} + \mu \right) }}{\pi d B_{og}/nH}, 
\label{eq:root1}
\end{equation}
and gives us a mean to explore the behaviour of the friction coefficient $\mu$ with the cohesion. \\[3pt]
\indent We first investigate the case of very weak cohesion, namely a Bond number approaching zero \bond$\rightarrow 0$. In that case, the quantity $ x = \frac{\pi d B_{og}}{nH} << 1$ becomes very small, and so does the term $y = 2x (\frac{x}{2} + \mu)$. Accordingly, equation (\ref{eq:root1}) can be expanded using  the expansion of $(1- y)^\alpha$ when $y\rightarrow 0$, thus giving:
\begin{equation}
X = \tan\alpha \simeq \mu + \frac12(1+\mu^2) \frac{\pi d}{nH} B_{og}.
\label{eq:root2}
\end{equation}
Setting  the coefficient of friction $\mu$ to the value $\mu = 1.24$ given by equation (\ref{eq:failure2}) and the data in the non-cohesive case \bond$=0$, and choosing $n=1$, we observe that the simplification (\ref{eq:root2}) operated for \bond$\rightarrow 0$ nicely describes the numerical data, displayed in Figure \ref{fig:prediction}. Interestingly, it gives a better description for \bond$=5$ than the complete solution (\ref{eq:root1}) does, using the same values  $\mu = 1.24$ and $n=1$, also shown if Figure \ref{fig:prediction}. We note that for these values of $\mu$ and $n$, equation (\ref{eq:failure2bis})  gives $B_c = 5.09.$\\[2pt]
\indent However, equation (\ref{eq:failure2bis}), which can be rewritten
\begin{equation}
\mu = \tan(\alpha) - \frac{ \pi d }{2 n H}  B_{og} (1+\tan^2\alpha),
\label{eq:failure3bis}
\end{equation}
discloses that $n\!=\!1$ imposes a negative value for the friction $\mu$ for Bond numbers larger than $\sim12$. The choice $n\!=\!1$ is thus not sustainable for large values of \bond. 
On the other hand, previous numerical work comparing  the stability of discrete cohesive piles with the stability of a continuum counterpart concluded that $n\simeq9 $ \cite{abramian20}.  In addition, equation (\ref{eq:failure3bis})  suggests that the friction $\mu$ may decrease with \bond\:  if the failure angle $\alpha$ does not compensate the variation of cohesion.  We thus  assume  $\mu= \mu_0 - a B_{og}$.  Setting $n=9$,  $\mu_0 = 1.73$ and $a = 0.0194$,   we plot the complete solution given by equation (\ref{eq:root1}).  We find that the prediction nicely describes the numerical data for $ 5 \leq B_{og}$, as shown in Figure \ref{fig:prediction}.\\[3pt]
 Our results have several implications. First, the successful predictions of the model coincides with the existence of a bifurcation in the behaviour of cohesive granular material, defining two distinct frictional behaviours. While weakly cohesive material exhibit a constant internal  friction, larger cohesion involve the significant decrease of friction with cohesive properties. Since stronger adhesive forces at contact between  grains signifies a more solid-like interface between the two sliding blocks,  and less erratic dissipative collisions in the shear band,  stronger cohesion inducing a smaller friction,  namely a cohesion-induced weakening, seems  a realistic scenario.   Our data show a cut-off value  between the two frictional regimes at a cohesion of $B_{og} = B_c \simeq 5$, as predicted by equation \ref{eq:failure3}. This cut-off value between the two regimes is compatible with the maximum cohesion probed by capillary forces, estimated to coincide with $B_{og} \simeq 4 $ by \cite{richefeu06}, for which friction is observed to be mainly constant. \\
 In addition, the computation of an equivalent continuum cohesive stress implies a different representative surface unit in both regimes: while weak cohesion may involve one grain diameter, stronger cohesion involves larger surfaces, {\em e.g.} $9$ grains diameters in our case, in agreement with a previous study on the discrete/continuum comparison \cite{abramian20}. This somewhat puzzling result might reflect a difference in stress transmission at contact in weakly and highly cohesive packings, and bespeak a different texture in both cases.  Further work is needed to explore this new line.

\section{Conclusion}

The failure of cohesive granular steps collapsing under gravity are simulated for a large range of cohesion. A first objective is  to establish a sensible criterion for capturing the failure characteristics. Focusing on the cumulative displacement of the grains, and defining a displacement threshold, we discuss a practical procedure to identify the signature of the failure and its robustness. We are able to locate the failure in time, showing that the different stages of the destabilisation are well-defined with small error bars. We find that the onset of the failure is delayed by increasing cohesion, while its duration becomes shorter.  Focussing on grains displacement we identify well-defined linear failure. Their orientation with the horizontal varies non-monotonously with cohesion, exhibiting first increasing, then decreasing angles  when cohesion increases.\\
   Solving the equilibrium of a cohesive column, we give evidence of the existence of a cut-off value of the cohesion between two distinct frictional behaviours. We are able to make successful predictions for the dependance between failure angle and cohesion, thereby disclosing the predicted two aforesaid behaviours: a constant  friction at small cohesion, and a significantly decreasing  friction with cohesive properties at larger cohesion.  Since stronger adhesive forces at contact between grains signify a more solid-like interface between two sliding blocks at failure,  and less erratic dissipative collisions, stronger cohesion resulting in smaller friction seems a sensible scenario. At any rate, this cohesion-induced weakening is apparent in our numerical cohesive failure process. \\[1pt]
 Although the choice for a specific displacement threshold is argued in the present paper, a quantitative study of how the choice of a threshold may change the picture of the failure, and more specifically its orientation, would be interesting. However, since the simplest equilibrium model, with friction and cohesion as sole ingredients, nicely describe the numerical data, without contradicting previous comparisons with continuum cohesive systems \cite{abramian20},   we are confident that the present conclusions would not be affected.

\section*{acknowledments}
This work is part of the COPRINT project (http: //coprint226940055.wordpress.com) supported by the ANR grant ANR-17-CE08-0017.

\end{document}